\documentclass[]{interact}

\def\blindIt{0} 

\usepackage{epstopdf}
\usepackage[caption=false]{subfig}

\usepackage[natbibapa,nodoi]{apacite}
 \usepackage{setspace} 

\usepackage{graphicx}
\usepackage{CJKutf8}
\usepackage[hyphens]{url} 
\usepackage{float} 
\usepackage{xfrac}
\usepackage{graphicx}
\usepackage{wrapfig, setspace, booktabs}
\usepackage{tabularx}
\usepackage{hyperref}
\hypersetup{
	colorlinks=true,
   	linkcolor=blue,
   	citecolor=blue,
    filecolor=blue,
   	urlcolor=blue}

\usepackage{booktabs}
\usepackage[caption=false]{subfig}

\newcommand{\urlDate}{last accessed 2019-09-07}
\newcommand{\furl}[1]{\footnote{\url{#1} (\urlDate).}}
\newcommand{\frurl}[1]{\footnote{\url{#1}}}

\usepackage[dvipsnames]{xcolor}
\definecolor{ao}{rgb}{0.0, 0.5, 0.0}

\usepackage{multirow}
\usepackage{tabularx}
\newcolumntype{L}[1]{>{\raggedright\arraybackslash}p{#1}}
\newcolumntype{C}[1]{>{\centering\arraybackslash}p{#1}}
\newcolumntype{R}[1]{>{\raggedleft\arraybackslash}p{#1}}

\usepackage{comment}
\expandafter\def\expandafter\UrlBreaks\expandafter{\UrlBreaks
  \do\a\do\b\do\c\do\d\do\e\do\f\do\g\do\h\do\i\do\j%
  \do\k\do\l\do\m\do\n\do\o\do\p\do\q\do\r\do\s\do\t%
  \do\u\do\v\do\w\do\x\do\y\do\z\do\A\do\B\do\C\do\D%
  \do\E\do\F\do\G\do\H\do\I\do\J\do\K\do\L\do\M\do\N%
  \do\O\do\P\do\Q\do\R\do\S\do\T\do\U\do\V\do\W\do\X%
  \do\Y\do\Z}

\usepackage{lineno}

\begin{document}


\newcommand{\mytitle}{Teaching Blockchain in K9-12: Instruction materials and their assessment}
\title{\mytitle}


\if\blindIt0
\author{ and Frank Breitinger}
\author{
\name{Leo Irudayam\textsuperscript{a} and 
Frank Breitinger\textsuperscript{a,b,*}
\thanks{\textsuperscript{*}The work was primarily done while being affiliated with University of Liechtenstein.}
}
\affil{ \textsuperscript{a}Institute of Information Systems, University of Liechtenstein, FL\\
\textsuperscript{b}School of Criminal Sciences, University of Lausanne, CH}
}
\fi
\if\blindIt1
\author{Blinded for review}
\fi

\maketitle

\begin{abstract}
This paper analyses the feasibility of including emerging IT-topics into secondary education (K9-12). We developed a class on Blockchain for K9-12 students which delivers relevant content but is also indented to motivate them to further engage with the topic. Our course consists of theory material as well as a hands-on application. Gamification techniques are utilised to encourage students and improve the learning experience.  The assessment, consisting of 166 students, shows a significant knowledge gain of students exposed to our materials especially when the theory and the hands-on materials are combined. Both, i.e., teaching material and chat application, are freely available. 

\end{abstract}


\section{Introduction}
Digitalisation and digitisation\footnote{According to \cite{gobble2018digitalization}, \textit{digitalisation} ``refers to the use of digital technology, and probably digitised information, to create and harvest value in new ways'' where \textit{digitisation}: ``is the straightforward process of converting analogue information to digital—turning pages into bytes.''}, 
buzzwords and strategic priorities of governments, public and private sector, cause significantly shorter adoption periods of emerging (IT) technologies. This, in return, results in a high demand of graduates filling these `digital jobs' which is a major challenge. 
A Market survey from 2018 in China by \cite{pwccn} has shown that the participants have ranked `talents' as the second biggest concern for Blockchain technology development.

One of the problems is that emerging topics are often only covered during higher education (e.g., first universities over degrees / classes in areas such as Blockchain, E-Sports, or Virtual Reality, have only recently been developed) and do not play a role in secondary education. 
We argue that this is too late as students 
(a) have already made their choice on what they will study while never being exposed to any emerging trends and 
(b) students not opting for higher education but a different career path like apprenticeships are not qualified.

Reasons for why emerging topics are not included earlier vary widely from teachers not having the technical expertise \citep[p. 551-552]{state_of_the_art}, being overloaded with work and cannot develop new materials \citep[p. 349-350]{Eickelmann2020}, or curricula only being not flexible (not allowing the integration of new topics). Consequently, students are pointed to other learning possibilities such as YouTube or Massive Open Online Courses (MOOCs). 

While the curriculum flexibility aspect is out of our hands, this article provides a free and open-source teaching activity on the emerging topic of Blockchain. In summary, the following contributions are made:

\begin{enumerate}
\item Theory (i.e., instructional, educational) material for a 45-60min lecture in English and German freely available for interested educators. While the content is already comprehensive, it may be adjusted as needed.
\item A hands-on experience utilising gamification to allow deepening the understanding of the domain. The hands-on experience is based on a self-developed Blockchain-Chat that is straightforward to set in any local network (GDPR compliant).
\item An assessment (pretest-posttest) of the developed materials clearly demonstrating that learning objectives were achieved and that students enjoyed the exercise (note: as the practical input requires additional time when utilising the content, we decided to evaluate it separately, demonstrating the importance / necessity of the practical element).
\end{enumerate}

The content of the teaching materials has been carefully chosen based on a brief analysis of current market demands. The topic Blockchain was used as (a) it is an omnipresent topic that applies to many different domains (cryptocurrencies, supply chain, e-government, etc.) and (b) it highlights one of the problems youth underestimates / neglects -- once something is posted on the Internet, it will always be there, and you do not have the control anymore. 

The remainder of this work is organised as follows: The next section highlights the \nameref{sec:relwork}. Sec.~\ref{sec:methodology} outlines our five-step methodology from how learning objectives were found to creating the materials. The core of this article are Sec.~\ref{sec:course_content} and Sec.~\ref{sec:assessment} describing the \nameref{sec:course_content} and the \nameref{sec:assessment}, respectively. 
In Sec.~\ref{sec:subjective_perception} we provide our subjective perception of delivering the course to students before we briefly outline some limitations in Sec.~\ref{sec:Limits}. The last section concludes this article.

\section{Background and related work}\label{sec:relwork}
The cognitive rigour (CR) matrix is the base to quantify knowledge levels and progresses. It is a blend of the cognitive domain taxonomy of \citet[p. 18]{Bloom1968} and the depth of knowledge model by \citet[p. 25]{webb1997criteria}. Bloom's model defines six levels reaching from `knowledge' towards `evaluation'. The model of Webb categorises it into four distinct groups, from recalling facts to extended thinking levels. In combination, they shape the matrix by \citet[p. 5]{Hess2009CognitiveRB}, which is the foundation to design and assess our course.

\subsection{Blockchain technology}
One of the current emerging IT topics is Blockchain \citep[p. 1-2]{Centobelli}. To justify its growing popularity, it is vital to know that the Blockchain proposition aligns with the general IT-security goals of confidentiality, integrity, and authenticity in an advanced way \citep[p. 126]{Hellwig2020}. The Blockchain itself is continuously maturing, having started with Blockchain 1.0, which was the impact in the financial industry, more specifically about emerging cryptocurrencies such as Bitcoin or Ripple - considered as alternatives to fiat currencies (\citealp[p. 14-15]{B10}, \citealp[p. 11]{Xu2019}). Blockchain 2.0 improved the idea by the upcoming potentials of smart contracts and decentralised applications \citep[p. 9]{Swan}. The future versions and their impacts are subject to scientific discussions. Mainly, smart contracts are expected to improve \citep[p. 11]{Xu2019}, the relevance in a truly digital society is expected to grow \citep[p. 118]{EFANOV2018116}, public sector adoption increases \citep[p. 65]{8338007}. Standardisation and interoperability are achieved \citep[p. 4]{Blockchain35}.

\subsection{Need for education / training}
\citet[p. 1528]{mohaghegh2016computational} state that tertiary education is not anymore the apt place to introduce computational concepts. Hence, emerging technology needs to be integrated into lower education. In contrast, first IT concepts like web development required about 13 years from growing popularity to integration into school curricula  (\citealp[p. 129-130]{Comparison}, \citealp[p. 3]{Israel}), the continuous rise and ongoing digitalisation have accelerated developments and adoptions by manifold. Therefore, it is plausible that the adoption time of future emerging topics will significantly decrease \citep[p. 79]{K12Worldwide}. With this, students' self-learning skills must substantially improve, and computational skills appear to be mandatory to assure quality learning. Private-public sector partnerships can accelerate the adoption of emerging technology, enrich the curricula and reinvigorate lower education \citep[p. 132-133]{Oyelere2019}.

Above all, the insufficient supply is mainly uncovered by the continuously growing demand for the Blockchain workforce. In an annual survey by \citet[p. 32-34]{deloitte}, 31 \% of the respondents claim the lack of skill within the organisation as a barrier to greater adoption, and 44 \% see access to talents as a severe obstacle. In a survey by \citet{gartnerbct}, 18 \% of participating CIOs have said that Blockchain skills are the hardest to find on the market. \citet[p. 26]{pwccn} have shown in their market survey 2018 in China that the participants have ranked `talents' as the second biggest concern for Blockchain technology development, above `industry standards' or `market readiness'.

Creating talents in a specific area are always limited to the number of students willing to pursue a career in this field. Hereby, STEM has its own characteristics. \citet[p. 17]{pbjl} postulate from their study that already one project-based STEM learning course can increase the interest in STEM in general. \citet[p. 25]{reinhold} enhance these findings, stating that any interactive task, application, hands-on activity or investigation contribute positively towards the students' interest. Although it appears relatively straightforward, \citet[p. 672-673]{stem_factors} still estimate the schools' influence as low and value social aspects like the parents' jobs as much more impactful. However, they request that schools take action to countersteer.

\subsection{Available materials and challenges}\label{sec:available_materials}
The Blockchain education landscape mainly consists of YouTube videos or other informal social platforms, MOOCs (Massive Open Online Courses), paid online courses and dedicated university programs. While YouTube videos are one of the most used sources \citep[p. 400]{Blockchain_from_where}, they are not designed to follow quality criteria, ensure unbiased information delivery, and mostly lack in providing a fundamental understanding to students of K9-12. MOOCs are a promising alternative but in comparison to paid online courses (starting around 100 dollars). However, \citet{Blockchaincourses} revealed a quality gap in his analysis stating only one free beginner's class makes it into the top-10 ranking. \citet[p. 2]{Firth2018TeachingBI} strengthens this concern saying that most online courses, including simulation or hands-on experience, are barely free. Dedicated university programs target advanced scholars and usually require preknowledge or expertise in a Blockchain application area.

Subsequently, there is a lack of high-quality, free Blockchain education material for K9-12 students. Additionally, there is almost no practical exercise in any course included. Combined with the aspect that open educational resources gain attention and attraction for the next generation of teachers \citep[p. 79-80]{usage_of_oer}, this observation demands greater importance.

German schools are recognised as under-average digitally developed institutions. In this regard, \citet[p. 240-241]{Bildungsberichterstattung2020} observed a weak technical infrastructure, a missing concept of teaching teachers to integrate digital solutions into their daily work and an unclarified legal situation, especially in terms of data privacy. Subsuming these aspects leads to a requirement that demands creating a privacy-friendly, easy-to-setup, self-explaining solution to achieve great acceptance at teachers.

\section{Methodology}\label{sec:methodology}
To identify, develop and assess our materials, we followed a five-step methodology: 

\begin{enumerate}
    \item Market analysis (results have already been discussed in Sec.~\ref{sec:available_materials}) to obtain an understanding of currently available education possibilities. 
    \item Defining learning objectives by utilising other existing courses as well as identifying most critical content by analysing job abs.
    \item Development of theory material.
    \item Development of hands-on material to deepen and apply the knowledge.
    \item Course assessment using three groups. 
\end{enumerate}

\noindent The subsequent paragraphs provide more details about each step.

\subsection{Market analysis} 
A common Internet search engine with the following keywords was used to identify existing courses: `Blockchain Course', `Blockchain 101', `Introduction Blockchain', `BCT' and `Bitcoin Course' and 'Ethereum Course'. We then manually analysed the twenty most prominent courses by their content coverage, duration, learning outcomes, demanded preknowledge, target group, and costs. The findings have been summarised in Sec.~\ref{sec:available_materials}.

\subsection{Defining learning objectives} 
The objective of the course is to provide students with the most relevant and essential aspects of Blockchain. Before developing a challenging and relevant class, specific learning and cognitive outcomes had to be defined. Therefore, learning objectives should be aligned to the demands of the Blockchain workforce. Consequently, we started by examining job ads found online in February 2021. In the next step, we translate these findings to learning objectives.

\paragraph{Job ads} 
To understand the most relevant Blockchain aspects, an analysis of job postings matching specific keywords from Stepstone, Indeed, and LinkedIn had been conducted. 
After cleanup, 397 distinct German and English postings from the German-speaking European countries, the US, Canada, and India matching the keywords `Blockchain', `DLT', `Bitcoin', `Ethereum', `Hyperledger' or `Tokenisation' remained. However, due to the misuse\footnote{Example: a job listing looking for people with `first practical experiences or interest in modern technologies or programming languages such as  Cloud-Services, BigData \& Data Analytics, AI, RPA, Blockchain or SQL, R, Python'}
of the terms, the qualification analysis has been conducted manually. The manually filtered postings have been examined by their demands regarding qualification and experience levels matching specific hard skills to the minimum class in the CR matrix.

\paragraph{Defined learning outcomes} These topics of the previous analysis have been placed on a CR matrix and can be positioned on upper levels or covered across the entire matrix. The last is ideal for an introductory course since they are perceived as the door-opener and most essential factors. Each of these base topics requires a more profound understanding of its basis itself. Hence, their preknowledge topics are collected and combined into a graph of dependencies for all important topics. With this, the HITS algorithm by \citet{HITS} had been used to weigh a topic by relevance for Blockchain technology and by relevance to justify other Blockchain topics.

\subsection{Theory material / teaching materials}
The planned introductory Blockchain course allocates 90 minutes for theory and practice ($M_A$) or 45-60 minutes for the theory-only group ($M_B$). Its layout follows the outcome-centred course design by \citet[p. 19-21]{nilson2016teaching}. As an outcome, the analysis has shown that a minimum requirement of hard skills is the `Understand' level (second lowest) in Bloom's taxonomy. Furthermore, quick wins are focusing on technical fundamentals, the general purpose of Blockchain technology and cryptocurrencies. On the other hand, specific (technical) applications or protocols demand a more profound understanding. 

\begin{figure}[ht]
\index{Design science research process (DSRP) model}
\begin{center}
\includegraphics[width=1.0\linewidth]{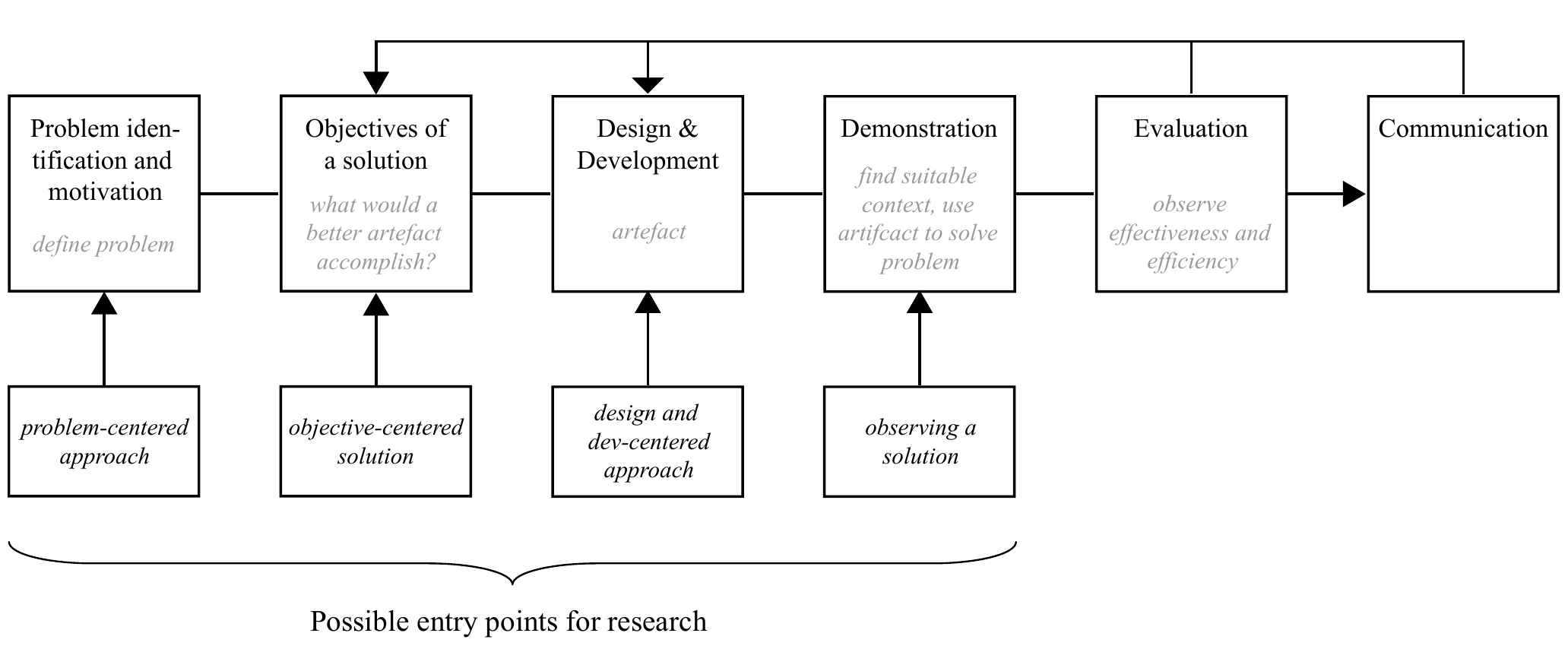}
\end{center}
\caption [Design science research process (DSRP) model]{Design science research process (DSRP) model (own figure, derived from \citet[p. 93]{peffers2007design}).}
\label{fig:dsrp}
\end{figure}

The outcome-centred course utilises the Design Science Research Process model by \citet[p. 93]{peffers2007design} visible in figure \ref{fig:dsrp}. It demands in advance a definition of the problem. For the course, this will be a fictive story highlighting a problem of today's world. Next, the objectives towards a solution are sketched. With this, the target artefact is designed and developed, heading to the demonstration. From here onwards, the solution artefact will be evaluated, and a message communicated.

\subsection{Development of hands-on material}
All previously derived aspects and requirements for creating hands-on materials have been joined and grouped in the specifications document. Non-functional requirements as the ease of setup, the platform independence and simplicity of the software have been the baseline to develop the application, which also allows less-technical instructors to use it (of course, an educator still needs familiarity with Blockchain to deliver the content). 
With these technical restrictions and guidance, the next step consisted of finding an ideal setup and use case for an application. Then, once the objectives and purpose of the application were defined, the actual development and implementation phases started.

First, mock-up prototypes have been designed and assessed by a small group of students from the upper secondary schools and early semester university students regarding usability, self-explainability, motivation and educational aspects. Second, a gamification layer has been used to enhance the students' motivation and guidance concept.

In the next step, potential technology stacks have been scouted and evaluated by the following factors: relevance of technology, distribution level, support by different platforms, compliance aspects and developer community. As a result of this, WebRTC-technology has been chosen for the application. Therefore, the following steps focus on creating an attractive, usable, and assessable minimal viable product under a free licence and open source.

\subsection{Assessment of materials}
The teaching materials were assessed by teaching them to students of Baden-W\"urttemberg (Germany) upper secondary school classes (general and specialised gymnasiums were included). Since school classes represent a fixed group of students, the field study was by default non-randomised. The resulting non-randomised control group pretest-posttest design was based on a distinguishment of an experimental group that received treatment (gets taught) and a control group that receives no treatment (needs to self-research - called $M_P$) \citep[p. 156]{Levy2011}. However, the course desired assessment by the value of the practical application. Hence, there were two independent treatment groups: one receiving theory and practical application ($M_A$) and the other theory only ($M_B$).

Students' previous knowledge shall be taken into the equation with the pretest to determine the success by the posttest scores harmonised by pretest scores. After the treatment, students will complete the same test as the posttest with additional opportunities to give feedback in the form of a 5-star rating and free-text comment.

The test consisted of twelve questions taking about 10-12 minutes to complete. Since more than 75\% of the students have answered the first pretest question correctly, it has been excluded from the analysis. The questions were grouped by Bloom taxonomy (more details are shown in table~\ref{tbl:test_design} in the appendix):
\begin{itemize}
    \item `Understand' (second-lowest level): total 21 points
    \item `Apply': total 14 pts
    \item `Analyse': total 7 pts
    \item `Evaluate': total 10 pts
\end{itemize}

In detail, three different statistical tests were conducted:
\begin{itemize}
	\item A t-test on the posttest score means of both treatment groups compared to the placebo group, to evaluate the success of teaching this subject compared to self-research homework.
	\item An F-Test ANCOVA analysis to evaluate which group scores best in the posttest (representing the success of the course) harmonised by the matching pretest result.
	\item A t-test on the students' grades to assess the dependency of performance in school to test result. Ideally, this test should show no significant correlation.
\end{itemize}

\section{Course content layout and development}\label{sec:course_content}
This section will first discuss the learning objectives, followed by the teaching materials, with an introduction to the hands-on materials. The assessment is covered in the next section.

\subsection{Derived learning objectives}
The job ad analysis served as a compass on which aspects under the umbrella term Blockchain are the most relevant. The following subtopic clusters have been defined:
\begin{itemize}
    \item Technical Blockchain aspects: consensus mechanisms, the security of networks, data storage, ...
    \item Protocol-specific subtopics: Ethereum, IOTA, Hyperledger, ...
    \item General understanding: Blockchain, Blockchain industry, tokenisation, ...
    \item Cryptocurrencies: generally, Bitcoin, Ethereum (Ether, ERC-Token), ...
    \item Blockchain applications: Solidity, dApp, decentralised identifiers, ...
\end{itemize}

These subtopic clusters were used on the data of the job ad analysis. In detail, the demanded requirements for a job posting have been matched with these subtopics and the knowledge/experience classification in the CR matrix. 
While different job levels (e.g., junior, senior, principal, etc.) significantly impacted the distribution of minimum levels of each subtopic, the most notable influencer remained the complexity and coverage of the aspect itself. 
Protocol-specific subtopics are very close to research and optimisation of de-facto standards. Therefore, they require the most in-depth understanding. General understandings can be seen as the most fundamental and equally important base and can be ideally used. 
The HITS algorithm allows to objectively weight subtopics by relevance (results visible in table~\ref{tbl:HITS} in the Appendix). 

\begin{figure}[H]
\index{Course contents grouped by learning objective in CR-Matrix.}
\begin{center}
\includegraphics[width=1.0\linewidth]{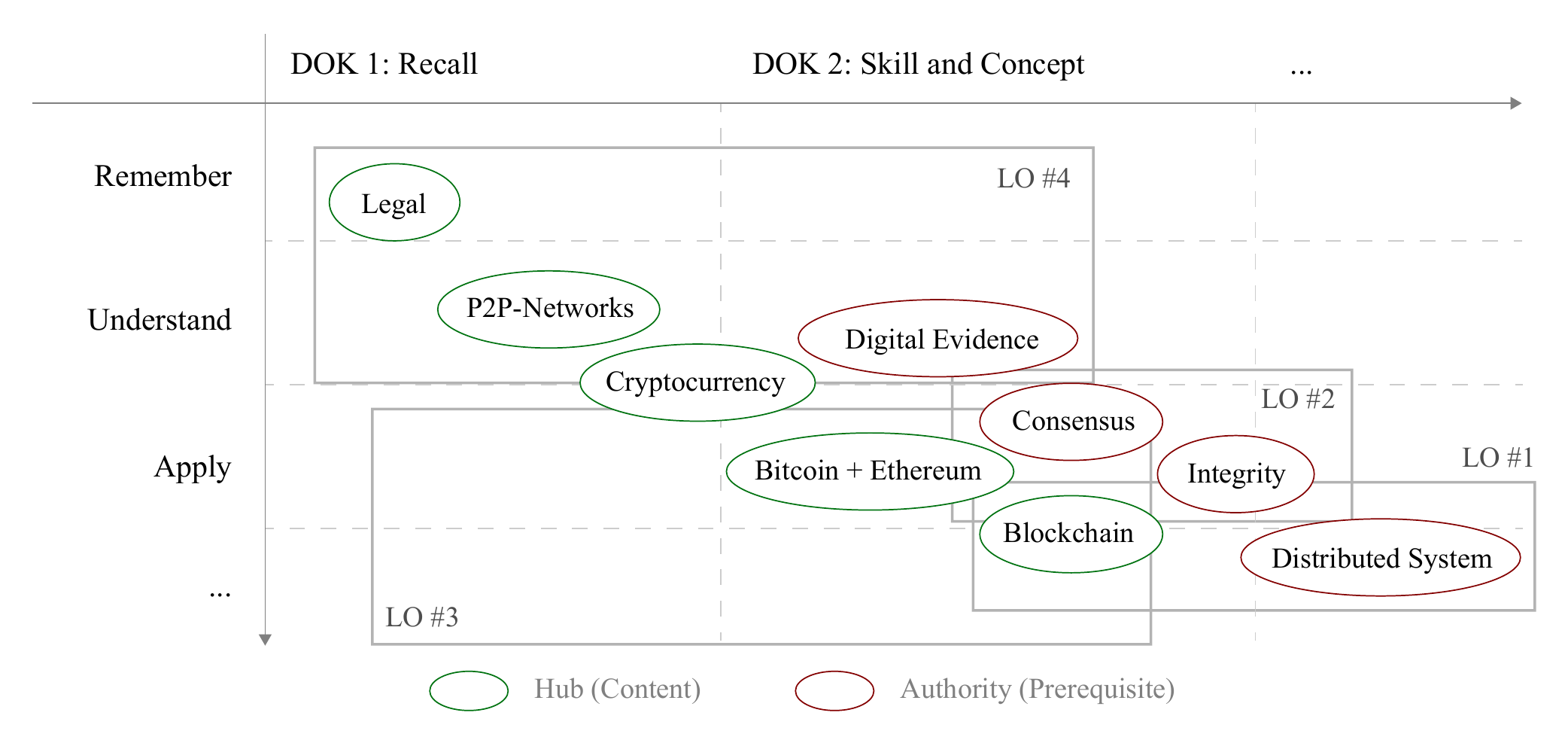}
\end{center}
\caption [Course contents grouped by learning objective in CR-Matrix.]{Course contents grouped by learning objective in CR-Matrix.}
\label{fig:cr_applied}
\end{figure}

The rankings (visualised in figure \ref{fig:cr_applied} in the CR-Matrix) lead directly to the following four learning objectives (LO) in the scheme by \citet{learning_objectives}:

\def\LOa{Students can explain the concept of distributed systems and can apply this knowledge to outline the architectural difference of Blockchain compared to client-server architectures}
\def\LOb{Students demonstrate the idea of integrity, tell what it means to achieve consensus and can practice this knowledge in the context of Blockchain technology}
\def\LOc{Students comprehensively present the basic concept of Blockchain technology and can locate use cases; ideally, some justify their decision}
\def\LOd{Students can name some of the (legal) implication Blockchain technology has}
\begin{itemize}
    \item LO \#1: \LOa.
    \item LO \#2: \LOb.
    \item LO \#3: \LOc.
    \item LO \#4: \LOd.
\end{itemize}

\subsection{Theory course design}
Joining the learning objectives of the previous subsection and the relevant preknowledge subtopics of the HITS algorithm led to the DSRP model adoption in the following course layout:

\begin{figure}[H]
\index{Theoretical Contents Framework}
\begin{center}
\includegraphics[width=1.0\linewidth]{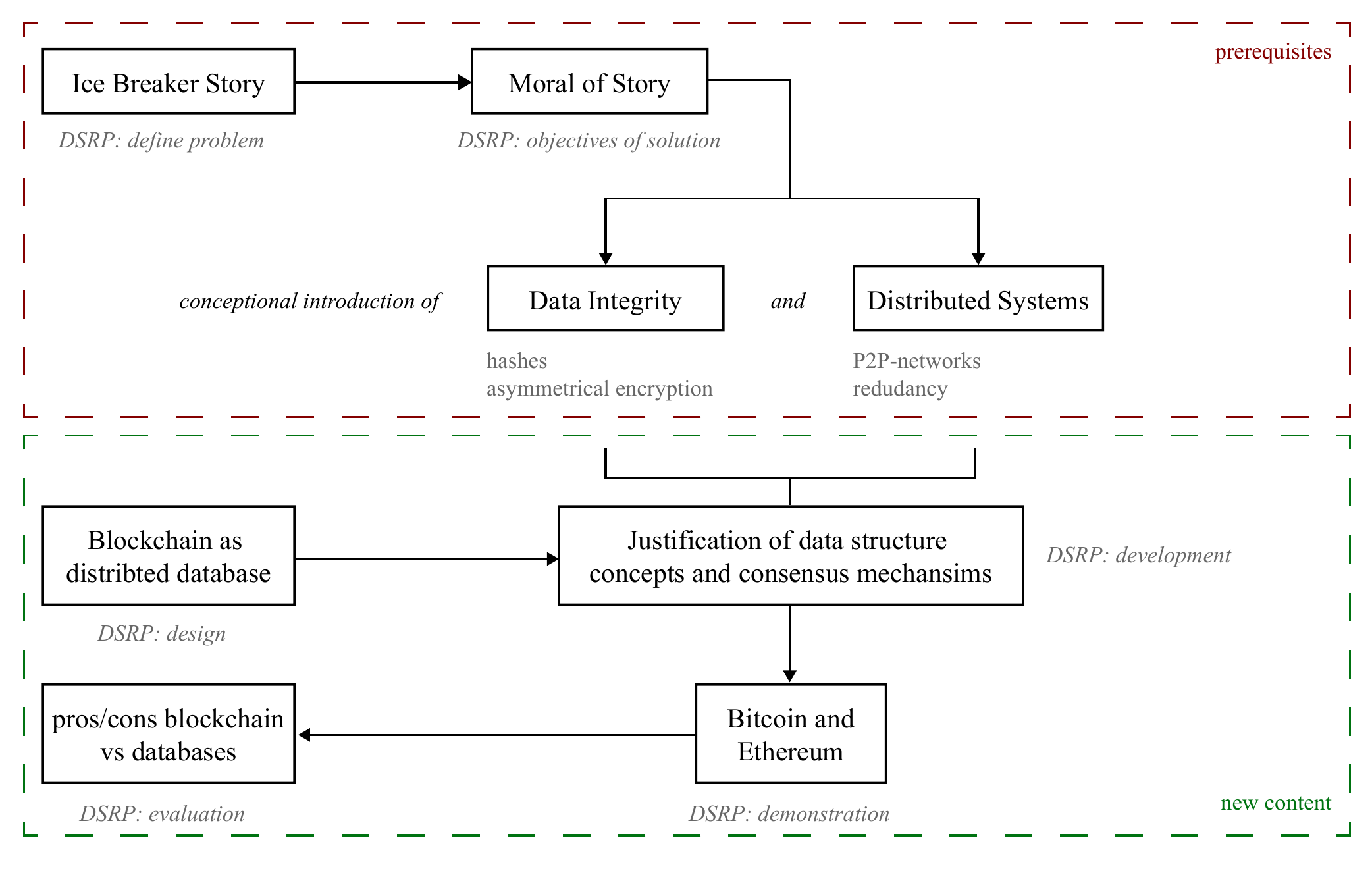}
\end{center}
\caption [Theoretical Contents Framework]{Theoretical Contents Framework (own figure)}
\label{fig:theoretical_contents}
\end{figure}

The ice breaker of figure~\ref{fig:theoretical_contents} is a story about the value of sensitive data and what a potential loss can cause (here: money on a bank account). This catchy and emotionalising ice breaker shall get students attached to the topic \citep[p. 48-49]{nilson2016teaching}. The moral is that a backup can restore data, but the end-user cannot verify the integrity of this data nor has a real piece of evidence for complaints. 

In the following steps, the prerequisites of the main topics are introduced and familiarised to lead to a Blockchain solution. In particular, the objectives are to create a network of peers sharing a single shared source of truth; data manipulation operations shall be transparently tracked and verifiable, and a transaction shall be verified and confirmed by an unrelated third party.

In the theory-only course, Bitcoin and Ethereum take the spot of demonstration objects. The theory and practice course is enhanced by the dedicated learning software simulating a Blockchain environment covered in a chat platform.

As a summing-up (evaluation and communication), the impact of Blockchain is discussed, mainly centring on its (legal) implications and how this technology might be utilised for preserving digital evidence. A quick recap, joined by a critical analysis of suitable use cases, concludes the lecture. In this regard, the environmentally unfriendly `meaningless' mining operation of the Proof-of-Work consensus mechanism will be criticised, and a glimpse of alternatives will be presented \citep[p. 1]{dotan2020proofs}.

This entire concept is verbalised within a PowerPoint slide deck purposefully incorporating a young adult design approach. It follows the mantra of a Reddit community `Explain Like I'm Five' of underlying, `how' and `why' phenomena exist simplistically without sacrificing content quality \citep[p. 2, p. 9]{fan2019eli5}. It is available for free usage in German and English on GitHub%
\if\blindIt1
\footnote{\textbf{Link blinded for review.}}.
\fi
\if\blindIt0
\frurl{https://github.com/lirudayam/chunkchain/tree/main/presentations}.
\fi

\subsection{Chat application}
The show-case application is a chat platform for students. Chat tools like WhatsApp or Signal are familiar to students and automatically invite them to interact with the platform. Compared to these free or proprietary solutions, the designed application enhances the theory course's educational aspect.

Students chat with each other based on a local Blockchain. Throughout their chat experience, they are requested to complete different missions to achieve the various levels unlocking more functionality. In the first missions, students rehearse some of the key facts from the theory input. Then, they must either read, answer a question, or interact with the software to complete a mission. Once completed, they are automatically navigated to the next mission. If they have completed the first missions, they are promoted to level 2.

\begin{figure}[ht]
\index{Screenshot of the interactive mining mission}
\begin{center}
\includegraphics[width=1.0\linewidth]{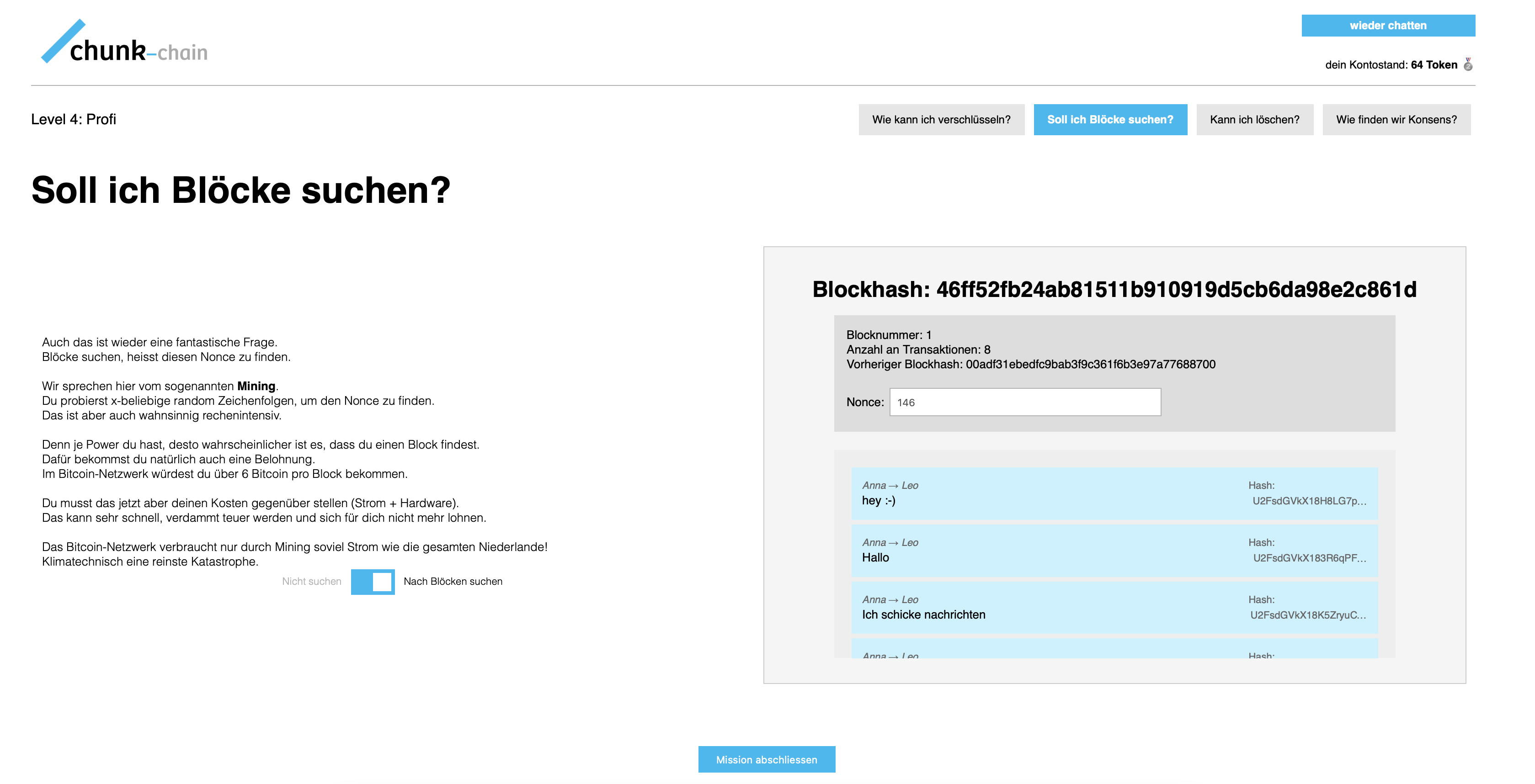}
\end{center}
\caption [Screenshot of the interactive mining mission]{Screenshot of the interactive mining mission (own figure). \textbf{Remark:} We are currently translating the application and in case the paper is accepted this screenshot will be replaced with the English version.}
\label{fig:screenshot_mining}
\end{figure}

From level 2 onwards, students get more in touch with the technical aspect of the software. For example, while they have previously heard how messages are stored in transactions and protected by hashes and encryption, they can see this in-depth for every transaction in the network. Furthermore, students can deep-dive into various aspects like data storage, the distributed network, how notifications are stored in blocks, how the chain evolves, how mining operations work (visible in figure~\ref{fig:screenshot_mining}), etc.

The entire application has been designed under the aspect of simplicity and benefit according to previously defined technical requirements. In this context, simplicity also covers the teachers' part. This software is equipped with a detailed README file containing all necessary setup steps. This application does not need any dedicated installation of other tools, any permissions by IT administrators or any support by trained staff. We observed a teacher who could set up the entire application in less than ten minutes without help.

Since it is designed as simple as possible, no dedicated hardware is required. Old computers with new browser versions support this application as well as new computers. With the local Blockchain aspect, there are no compliance issues. No data is stored outside the network, and the information automatically erases when all participants close their browser window. In short, this application is entirely GDPR-compliant and privacy-friendly.

Although a chat platform does not represent a practical Blockchain application, its gamified and social approach raise motivation and curiosity in the students - ideally suited for this generation's learners (\citealp[p. 4]{kiryakova2014gamification}, \citealp[p. 18]{boiko}). Moreover, within the field study, the expected engagement and satisfaction rates have been entirely met, confirming the software's value proposition.

It is available for free usage in German and English on GitHub
\if\blindIt0
\furl{https://github.com/lirudayam/chunkchain}
\fi
\if\blindIt1
\footnote{\textbf{Link blinded for review.}}.
\fi

\section{Assessment}\label{sec:assessment}

The materials have been evaluated in a field study, including 166 students who have completed pretest and posttest. 
Teachers had been given free choice to select a group ($M_A, M_B, M_p$) and consequently selected the group with the best-expected learning experience for their students and available in-class time.
Table~\ref{tbl:distribution_participants} reveals that most students of this assessment were assigned to group $A$. 
Remark: Since the field study took place around the A-levels, teachers of the final A-level grade often declined to participate.

\begin{table}[ht]
\centering
\index{Distribution of participants by cohort.}
\begin{tabular}{lp{2cm}p{2cm}p{2cm}|l}
& $M_A$ & $M_B$ & $M_P$ & $\sum$ \\ 
\midrule
Last A-Level grade     & 12          & 0           & 0       & 12        \\
Prelast A-Level grade  & 28          & 42           & 13       & 83        \\
3rd last A-Level grade & 40          & 9           & 22       & 71       \\
\midrule 
$\sum$ & 80 & 51 & 35 & 166 \\
\end{tabular}
\caption{Distribution of participants by cohort.}\label{tbl:distribution_participants}
\end{table}

\subsection{Descriptive statistics}

\begin{figure}[th]
\index{Distribution of pretest and posttest scores}
\begin{center}
\includegraphics[width=1.0\linewidth]{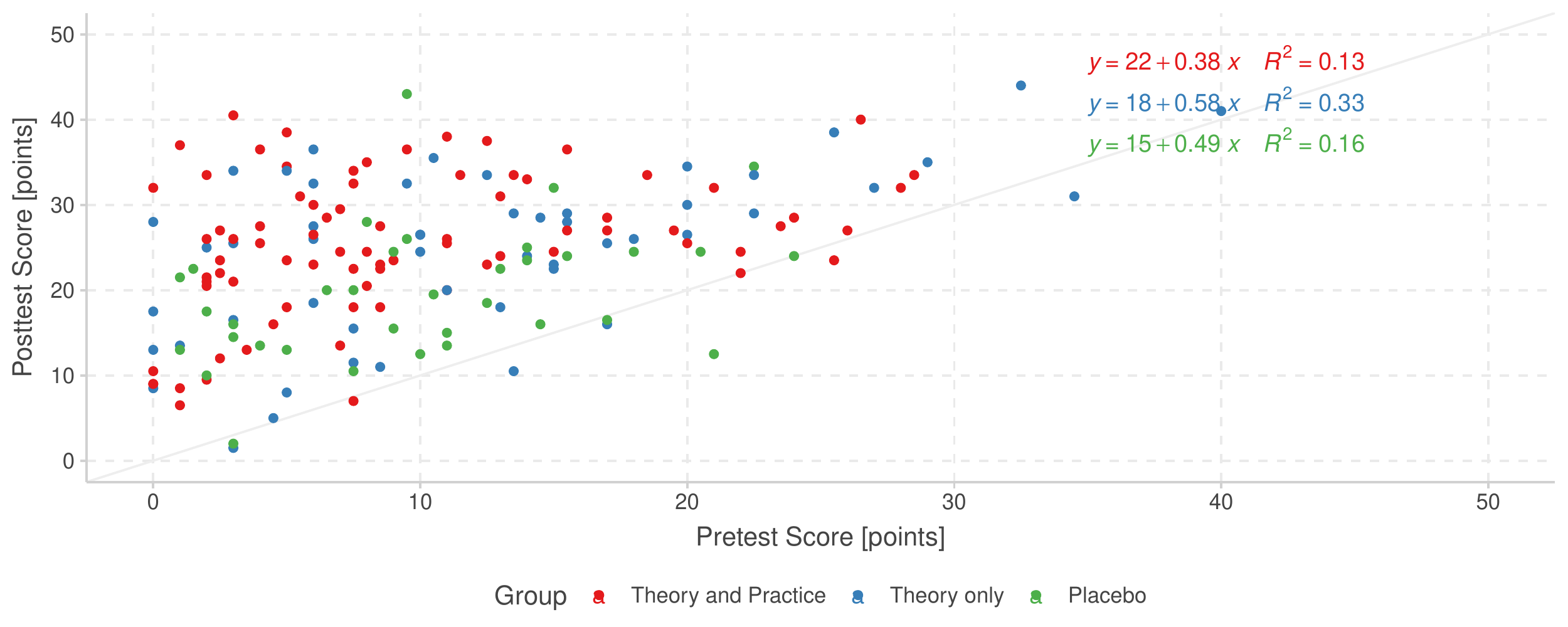}
\end{center}
\caption{Distribution of pretest and posttest scores by treatment group.}
\label{fig:distribution_tests}
\end{figure}

For the assessment, posttests were linked to matching pretests. Figure~\ref{fig:distribution_tests} visualises the relation of these tests. It attests that a low pretest score does not imply a weaker posttest score. Furthermore, a visible majority has improved in the test. Outliers are mostly the result of incomplete posttest submissions.

These results can be analysed more granular in the perspective of applying Bloom's taxonomy (table~\ref{tbl:distribution_testscores_cr}). In absolute terms, each viewed class of Bloom's taxonomy has shown an increase independent of the group on average. However, $M_A$ is always best followed by $M_B$ and $M_P$. The same view can be applied by learning objective (table~\ref{tbl:distribution_testscores_lo}) to verify the success by subject areas.

\begin{table}[th]
\centering \small
\begin{tabular}{llllll}
           &                        & Understand (21pt) & Apply (14pt) & Analyse (7pt) & Evaluate (10pt)  \\ 
\midrule
\multirow{3}{*}{Pretest $Y_1$}    & $M_A$                 & 4.61    & 1.45    & 1.86        & 1.56                   \\
           & $M_B$  & 5.72    & 1.61    & 2.56        & 2.47              \\
           & $M_P$  & 4.66    & 1.14    & 2.46  & 1.91              \\ 
\midrule
\multirow{3}{*}{Posttest $Y_2$}  &  $M_A$                 & 10.54 \textcolor{Green}{(+ 5.92)}     & 5.80 \textcolor{Green}{(+ 4.35)}   & 3.88 \textcolor{Green}{(+ 2.01)}        & 5.36 \textcolor{Green}{(+ 3.81)}              \\
           & $M_B$  & 10.15 \textcolor{Green}{(+ 4.44)}   &  5.39 \textcolor{Green}{(+ 3.78)}   & 3.62 \textcolor{Green}{(+ 1.04)}  & 5.59 \textcolor{Green}{(+ 3.12)}       \\
           & $M_P$ & 7.67 \, \textcolor{Green}{(+ 3.01)}   & 4.71 \textcolor{Green}{(+ 3.57)}    & 2.91 \textcolor{Green}{(+ 0.46)}       & 4.40 \textcolor{Green}{(+ 2.49)}  \\ 
\end{tabular}
\caption{Mean scores by Bloom's taxonomy level (rounded by two digits).}
\label{tbl:distribution_testscores_cr}
\end{table}
           
\begin{table}[th]
\centering \small
\begin{tabular}{llllll}
           &                        & LO1 (13pt) & LO2 (12pt) & LO3 (32pt) & LO4 (11pt)  \\ 
\midrule
\multirow{3}{*}{Pretest $Y_1$}    & $M_A$                 & 3.35    & 2.30    & 4.43        & 1.09                   \\
           & $M_B$  & 3.83    & 2.77    & 6.33        & 1.43              \\
           & $M_P$  & 3.11    & 1.26    & 5.06  & 1.09              \\ 
\midrule
\multirow{3}{*}{Posttest $Y_2$}  &  $M_A$                 & 8.39 \textcolor{Green}{(+ 5.04)}     & 5.55 \textcolor{Green}{(+ 3.25)}   & 15.80 \textcolor{Green}{(+ 11.37)}        & 4.09 \textcolor{Green}{(+ 3.00)}              \\
           & $M_B$  & 8.08 \textcolor{Green}{(+ 4.25)}   &  5.08 \textcolor{Green}{(+ 2.31)}   & 15.25 \textcolor{Green}{(+ 8.92)}  & 4.18 \textcolor{Green}{(+ 2.75)}       \\
           & $M_P$ & 6.10 \textcolor{Green}{(+ 2.99)}   & 4.30 \textcolor{Green}{(+ 3.04)}    & 12.11 \textcolor{Green}{(+ 7.05)}       & 3.05 \textcolor{Green}{(+ 1.96)}  \\ 
\end{tabular}
\caption{Mean scores by Learning Objective (rounded by two digits).}
\label{tbl:distribution_testscores_lo}
\end{table}
           
Furthermore, students had the option (voluntary) to rate the course on a 5-star scale (5 being highest/best; table~\ref{tbl:ratings}) and could give feedback in free-text form. In summary, the students have very well received the overall course and the practical application (the lower practical ratings are primarily due to connectivity issues). Throughout the study, the author's impression had been even strengthened by the numerous expressions of gratefulness and recognition of relevance from many students and teachers. 
With this, the content as well as the course design, presentation, interactivity, and relevance, have been highlighted.

\begin{table}[ht]
\centering
\begin{tabular}{lllllll}
           &                        & Theory & Practical & N Samples  \\ 
\midrule
\multirow{4}{*}{$M_A$}    & Last A-Level grade      &  5.00  & 4.41        & 12       &              \\
           & Prelast A-Level grade  & 4.52    & 4.43    & 28        &        &        \\
           & 3rd last A-Level grade & 4.39    & 4.34    & 40        &        &  \\ 
\cline{2-7}
           & Total                 & 4.55    & 4.39   & 80        &        &   \\
\midrule
\multirow{4}{*}{$M_B$}   & Last A-Level grade     & -    & -     &         &        &    \\
           & Prelast A-Level grade  & 4.59    & -    & 42        &        &   \\
           & 3rd last A-Level grade & 4.50    & -    & 9        &        &  \\ 
\cline{2-7}
           & Total                 & 4.58    & -    & 51        &        &  \\
\end{tabular}
\caption{Mean ratings for the course (5-point star rating scale).}
\label{tbl:ratings}
\end{table}

\subsection{Inferential statistics}          
In the following the data presented in a descriptive way will be used for inferential statistics to address the statistical tests. 
The first test (all statistical prerequisites met) to evaluate the success of the course framework in comparison to self-learning (group $M_p$) has unveiled that the null hypothesis, postulating that students taught with the course materials have performed better than students researching on their own, can be rejected (p = 0.0027 $<$ 0.05). By grade level, this shows that not every grade level can confirm this:

\begin{table}[th]
\centering
\begin{tabular}{llllllll}
                                  & $df$ & $t$ & $p$ & $p < 0.05$  \\ 
\midrule
    Last A-Level grade     & -    & -    & - & -   \\
    Prelast A-Level grade  & 81    & -2.1357     & 0.0357 & *    \\
    3rd last A-Level grade & 69    & -1.3175     & 0.1921 &     \\ 
\end{tabular}
\caption{t-test for learning framework effectiveness by grade level.}
\label{tbl:t_test}
\end{table}

\begin{figure}[ht]
\begin{center}
\includegraphics[width=1.0\linewidth]{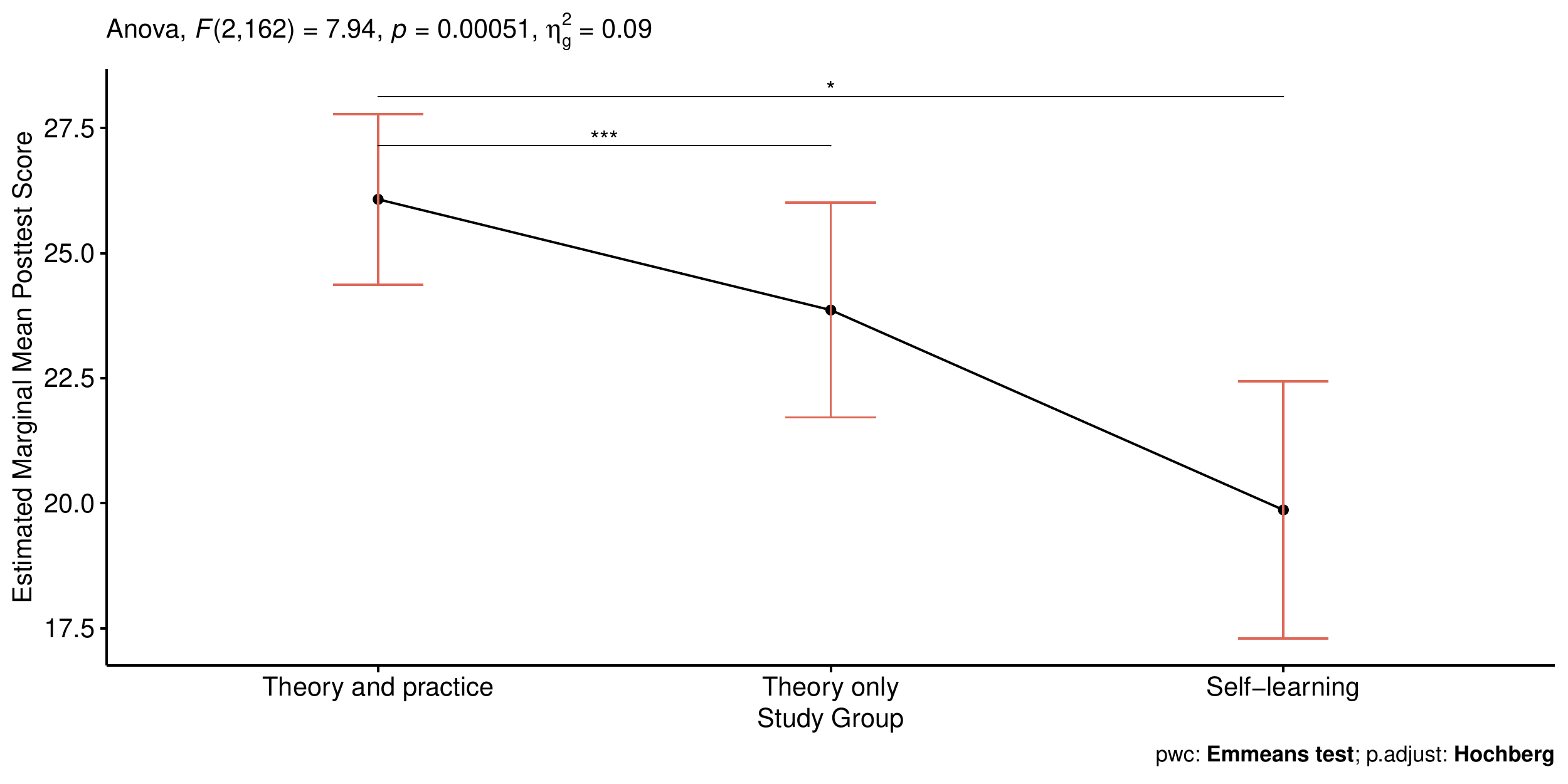}
\end{center}
\caption{Estimated marginal means of posttest scores by treatment group.}
\label{fig:ancova_menas}
\end{figure}

The second statistical test - ANCOVA analysis - has been conducted to prove difference by treatment. Figure~\ref{fig:ancova_menas} shows that the group ($M_A$) receiving theoretical and practical input has a significantly greater adjusted posttest mean score than group $M_B$. The posttest score means are adjusted by the pretests (covariate) and use an Estimated Marginal Means to illustrate the true treatment impact \citep[p. 217]{Searle1980}.

This concludes that the null hypothesis can be rejected since $p < 0.05$, stating that the score gains significantly differ by treatment. Subsequently, each treatment group substantially differs in the outcome (harmonised posttest score). Students of the theory and practice group have, as expected, the highest posttest scores, indicating they have learned the most compared to theory only and self-learning students (visible in figure \ref{fig:ancova_menas}).

The final hypothesis of investigating a linkage of posttest score towards overall school performance led to following results (one entry had been removed since it was an outlier):
\begin{gather*}
df = 108\\
t = -1.7275\\
p = 0.0869\\
cor = -0.1640
\end{gather*}

Even the potentially low correlation cannot be confirmed since $p > 0.05$, not allowing to reject the null hypothesis. Subsequently, a correlation between posttest scores and school performances cannot be assumed.

\section{Subjective perception}\label{sec:subjective_perception}
Within the field study, the authors have been confronted with different students with different backgrounds. To our astonishment, we have noticed that all included students have actively participated in the course. Students have raised numerous questions during and after the lecture, primarily by linking existing knowledge to received inputs. The gamification aspect of the practical application has visibly contributed to a productive learning environment. Students have encouraged each other to achieve the succeeding levels and actively helped each other when struggling.
The feedback, besides the rating scales, has been positive. The topic's relevance, course design, practical application, integration into the curricula, young-adult focused scheme, and interactive mode has been valued. Negative comments were exclusively addressing issues of remote teaching technology.
Teachers have valued the course as very helpful and well structured. They have acknowledged the topic's relevance and have reported a strong perception by the students. The easy setup and the freely available software appear to satisfy the demands. Teachers have said that it is straightforward to set up and easy to use compared to other tools. Throughout the comments have been very positive, enhanced by ideas of teachers to improve the course.

\section{Limitations}\label{sec:Limits}
The assessment of the essays was performed manually, which means that human error might have been introduced, e.g., at assessing open-ended questions. Furthermore, we encountered a few missing posttests, which meant that the matching pretests results had been excluded as well. Since the tests have been conducted unobserved, students could have cheated in either test.
However, due to the large number of complete test pairs that we could assess, we believe that our analysis is representative in both breadth and depth.

\section{Conclusion}\label{sec:conlusion}
Emerging (IT) technologies such as Blockchain provide the unforeseen potential for companies as well as governments in various ways. Hence, there is a high demand for talented works resulting in a significant workforce gap that higher education alone cannot close. In this study, we assessed the feasibility to included one emerging IT topic -- Blockchain -- in secondary education. We, therefore, analysed current market needs by reviewing job advertisements from various platforms leading us to four learning objectives (LO) for this introductory class. 
Based on the LO, we first developed theory materials (i.e., teaching materials) followed by a Blockchain chat application which is both freely available. The application has been developed so that it can easily be used by anyone and is by design GDPR compliant (i.e., the data students exchange remains in the local network and will be deleted after the simulation ends).
The assessment showed that despite the complexity of the topic, students were motivated and successful in enhancing their knowledge in this domain; treatment groups obtained higher scores.

\section*{Compliance with Ethical Standards}
According to the Office for Human Research Protections this Work is exempt: ``Research, conducted in established or commonly accepted educational settings, that specifically involves normal educational practices that are not likely to adversely impact students’ opportunity to learn required educational content or the assessment of educators who provide instruction. This includes most research on regular and special education instructional strategies, and research on the effectiveness of or the comparison among instructional techniques, curricula, or classroom management methods.''

All procedures performed in studies involving human participants were in accordance with the ethical standards of the Human Research Ethics Committee (HREC) and with the 1964 Helsinki declaration and its later amendments or comparable ethical standards.

\bibliographystyle{apacite}
\bibliography{references.bib}

\appendix
\renewcommand{\thesubsection}{\Alph{subsection}}
\section{Tables}

\begin{table}[H]
\index{Pretest/Posttest Design}
\centering
\begin{tabular}{lrll}
 & \multicolumn{1}{l}{Points} & Objective       & Task                                                                \\ 
(Q1)       & (2)    & LO \#4                                                            & (Match Bitcoin and Ethereum the umbrella term)   \\ 
\hline
Q2       & 6    & LO \#1                                                             & Map the terms to either Blockchain or DB       \\ 
\hline
Q3       & 3    & \begin{tabular}[c]{@{}l@{}}LO \#1 \\LO \#2\end{tabular}                                                             & Clarify which attributes distributes systems have                   \\ 
\hline
Q4       & 5    & LO \#2                                                             & Outline different options for ensuring data integrity               \\ 
\hline
Q5       & 2    & LO \#3            & \begin{tabular}[c]{@{}l@{}}Clarify what it means to achieve consensus\\ in a Blockchain network\end{tabular}  \\ 
\hline
Q6       & 6    & LO \#4            & Distinguish Bitcoin from Euro                                       \\ 
\hline
Q7       & 4    & LO \#3                                                             & Illustrate how the chain evolves                                    \\ 
\hline
Q8       & 4    & \begin{tabular}[c]{@{}l@{}}LO \#1 \\LO \#2 \\LO \#3\end{tabular}           & \begin{tabular}[c]{@{}l@{}}Distinguish the data structure and \\ storage methodology for both\end{tabular}    \\ 
\hline
Q9       & 5    & \begin{tabular}[c]{@{}l@{}}LO \#3 \\LO \#4  \end{tabular} & Clarify disadvantages of Blockchains                                \\ 
\hline
Q10      & 7    & LO \#3                                                               & Map use cases to technologies                                       \\ 
\hline
Q11      & 2 + 3  & LO \#3                                                               & Justify the decision for supply chain management system             \\ 
\hline
Q12      & 2 + 3  & LO \#3                                                             & Justify the decision for production management system               \\ 
$\sum$         & 54                         & all LOs                                                                   &                                                           
\end{tabular}
\caption [{Pretest/Posttest Design}]{Pretest/Posttest Design}
\label{tbl:test_design}
\end{table}

\begin{table}[H]
\index{Subset of results of topic valuation with HITS}
\centering
\begin{tabular}{ll|ll}
\textbf{Hubs (Content)} & \multicolumn{1}{l}{\textbf{Score ($h_p$)}} & \textbf{Authority (Prerequisite)} & \textbf{Score ($a_p$)}  \\ 
\midrule
Blockchain              & 0.140966528                         & Distributed System                & 0.174287458      \\
Bitcoin                 & 0.13350727                          & Integrity                         & 0.132040876      \\
Ethereum                & 0.099211033                         & Consensus                         & 0.113485709      \\
Cryptocurrency          & 0.082029696                         & Blockchain                        & 0.077000145      \\
P2P Networks            & 0.071036191                         & Cryptocurrency                    & 0.077000145      \\
Legal                   & 0.068691903                         & Digital Evidence                  & 0.070006124      \\
Redundancy              & 0.068515189                         & Single Point of Truth             & 0.053590137      \\
Fault Tolerance         & 0.056099092                         & Transactions                      & 0.046641983      \\
Transactions            & 0.053503432                         & Network Architectures             & 0.031390826      \\
Hashes                  & 0.039903748                         & Availability                      & 0.030556695          
\end{tabular}
\caption [{Subset of results of topic valuation with HITS}]{Subset of results of topic valuation with HITS}
\label{tbl:HITS}
\end{table}

\end{document}